\documentclass{PoS}

\usepackage{xspace}
\usepackage[numbers,sort&compress]{natbib}
\usepackage{caption}
\usepackage{floatrow}
\usepackage{varwidth}
\usepackage{amsmath}
\usepackage{amsbsy,amssymb}

\newcommand{\todo}[1]{}

\newfloatcommand{capbtabbox}{table}[][0.49\textwidth]

\newcommand{\heft}{HEFT\xspace}
\newcommand{\mh}{\ensuremath{m_{\mathrm{H}}}\xspace}
\newcommand{\mt}{\ensuremath{m_{\mathrm{T}}}\xspace}
\newcommand{\pth}{\ensuremath{p_{t,\mathrm{H}}}\xspace}
\newcommand{\Higgs}{\ensuremath{\mathrm{H}}\xspace}
\newcommand{\Hj}{\ensuremath{\Higgs\!+\! jet\xspace}}
\newcommand{\hj}{\Hj\xspace} 
\def\POWHEG{{\sc POWHEG-BOX-V2}\xspace}
\def\Gosam{{\sc GoSam}\xspace}
\def\Ninja{{\sc Ninja}\xspace}
\def\OneLoop{{\sc OneLoop}\xspace}
\newcommand{\ftap}{FT$_{\mathrm{approx}}$\xspace}
\newcommand{\CC}{C++\xspace}

\title{Top quark mass effects in HJ production at NLO}
\ShortTitle{Top quark mass effects in HJ production at NLO}

\author{\speaker{Matthias Kerner}
\\
Max Planck Institute for Physics, F\"ohringer Ring 6, 80805 M\"unchen, Germany\\
        E-mail: \email{kerner@mpp.mpg.de}}

\abstract{
  We present predictions for gluon-fusion Higgs boson production in association with one jet  at next-to-leading order in QCD. The calculation is performed retaining the full dependence on the top-quark mass. The two-loop integrals appearing in the virtual corrections are calculated numerically using the program \textsc{SecDec}. We show how a change of master integrals can improve the stability and runtime of the calculation.
We study the Higgs boson transverse momentum distribution and compare our predictions with approximated results.
}

\FullConference{Loops and Legs in Quantum Field Theory (LL2018)\\
		29 April 2018 - 04 May 2018\\
		St. Goar, Germany}

\begin{document}

\section{Introduction}
One of the major goals of the physics program at the LHC is the detailed study of the properties of the Higgs boson. To achieve this, accurate theoretical predictions of the various production and decay channels of  the Higgs boson are required. For the dominant production mechanism, gluon fusion via a top-quark loop, most theoretical predictions are obtained in an effective field theory approach, called \heft in the following, where the top-quark loop is integrated out by considering the limit $m_t\rightarrow \infty$. However, this approximation is only valid if the top-quark mass is much larger than all other scales of the process. While this is the case for inclusive Higgs production, one can expect that the top-quark mass effects become more important for the transverse momentum distribution of the Higgs boson, when $\pth \gtrsim \mt$.

Here we present the results of Ref.~\cite{Jones:2018hbb}, where the next-to-leading~order~(NLO)  QCD predictions for \hj production retaining the full top-quark mass dependence have been calculated. While this process is known to NNLO in the \heft approach~\cite{Boughezal:2013uia,Chen:2014gva,Boughezal:2015dra,Boughezal:2015aha}, beyond the leading order~\cite{Ellis:1987xu,Baur:1989cm} only approximated results of the top-quark mass effects have been known previously. To improve the \heft predictions, expansions in $1/\mt^2$ have been calculated~\cite{Harlander:2012hf,Neumann:2014nha} and combined with the exact Born and real radiation contributions~\cite{Neumann:2016dny}.  Recently also an expansion valid in the high-\pth region, where \mh and \mt are considered small, has been computed~\cite{Kudashkin:2017skd} and combined with the exact Born and real radiation~\cite{Lindert:2018iug,Neumann:2018bsx}, see also the presentation in Ref.~\cite{KudashkinLL}. Furthermore, contributions due to a bottom quark-loop are known~\cite{Mueller:2015lrx,Melnikov:2016qoc,Melnikov:2017pgf,Lindert:2017pky}.

The most challenging part of calculating the full NLO corrections for \hj production are the virtual contributions, which involve two-loop four-point integrals with internal masses. So far, only the planar integrals are known analytically~\cite{Bonciani:2016qxi} and progress on the calculation of the non-planar integrals is being presented in Ref.~\cite{FrellesvigLL}. In our calculation we obtain the virtual amplitude from  evaluating all loop integrals numerically. In the following we summarize our computational method and discuss how a change of the master integral basis can improve the numerical accuracy and run time of the virtual amplitude. We then present phenomenological results and compare them to approximated predictions.

\section{Details of the Calculation}
\subsection{ Computational Method}
In our calculation of the NLO corrections to \hj production, we have separated the implementation of the virtual corrections from all other parts, which we implemented in the \POWHEG framework~\cite{Alioli:2010xd}, taking advantage of the existing \hj NLO calculation in the HEFT approach~\cite{Campbell:2010cz,Campbell:2012am}. The LO matrix elements with full top-mass dependence are based on Ref.~\cite{Baur:1989cm}, and the real radiation contributions have been generated with \Gosam~\cite{Cullen:2011ac,Cullen:2014yla}. The libraries \Ninja~\cite{Mastrolia:2012bu,vanDeurzen:2013saa,Peraro:2014cba} and \OneLoop~\cite{vanHameren:2010cp} are used to perform the tensor reduction and for evaluating the scalar one-loop integrals, respectively. The contribution of the virtual corrections, described in the following, are added to the other contributions at the histogram level.

The implementation of the virtual contributions closely follows the method of Refs.~\cite{Borowka:2016ehy,Borowka:2016ypz}. We generate all two-loop diagrams with the program \textsc{Qgraf}~\cite{Nogueira:1991ex} and we express the amplitude in terms of a form-factor decomposition. Using a customized version of \textsc{Reduze}~\cite{vonManteuffel:2012np}, we obtained the full reduction of all contributing loop integrals to a set of master integrals, retaining the full dependence on $s,\,t,\,m_T,$ and $m_H$. However, the total size of the reduction files is about $1\,\mathrm{TB}$ and we therefore chose to use a previously obtained version of the reduction, where the mass ratio $m_H^2/m_T^2 = 12/23$ has been fixed. 
Using this simplified version of the reduction, we write the amplitude in terms of a 
quasi-finite basis of master integrals~\cite{vonManteuffel:2014qoa}, which is advantageous for the numerical evaluation of the loop integrals. The result presented in Ref.~\cite{Jones:2018hbb} have been obtained using a finite basis, where integrals with a small number of dimension shifts and squared propagators has been chosen. In section~\ref{sec:improvements} we describe how the choice of a master integral basis can be further improved.

The master integrals are calculated numerically with the program \textsc{SecDec}~\cite{Borowka:2015mxa,Borowka:2017idc}. For the numerical integrations, a quasi-Monte Carlo~(QMC) algorithm~\cite{Li:2015foa,QMCActaNumerica,nuyens2006fast} is used, where the worst-case error scales as $\mathcal O(1/n)$ with the number of sampling points $n$ for sufficiently smooth integrand functions.
The amplitude is implemented in the form of a \CC code which utilizes the Boost multiprecision library to guarantee a numerically stable evaluation of the coefficients. The integration of the loop integrals is parallelized using GPUs\footnote{See Ref.~\cite{JonesLL2016} for more details.} and we use the algorithm described in Ref.~\cite{KernerLL2016} to set the number of sampling points for each integral  such  that the total run time for obtaining each form factor with a precision of 0.5\% is minimized.

In the next section we describe how the stability and run time of our code has been improved since the publication of Ref.~\cite{Jones:2018hbb}. However, the calculation of the virtual amplitude still requires more then $1h$ computing time per phase-space point.
While this method allows us to obtain the fully differential NLO result, we also plan to include our results in a grid interpolation framework, which will then allow for a fast evaluation of the virtual contributions. This grid interpolation,  which will also be included in the {\sc{POWHEG-BOX}} framework, will then allow for further phenomenological studies beyond the fixed order.

\subsection{Change of master integral basis}
\label{sec:improvements}

\begin{figure}[t!]
  \centering
   \includegraphics[width=0.49\textwidth]{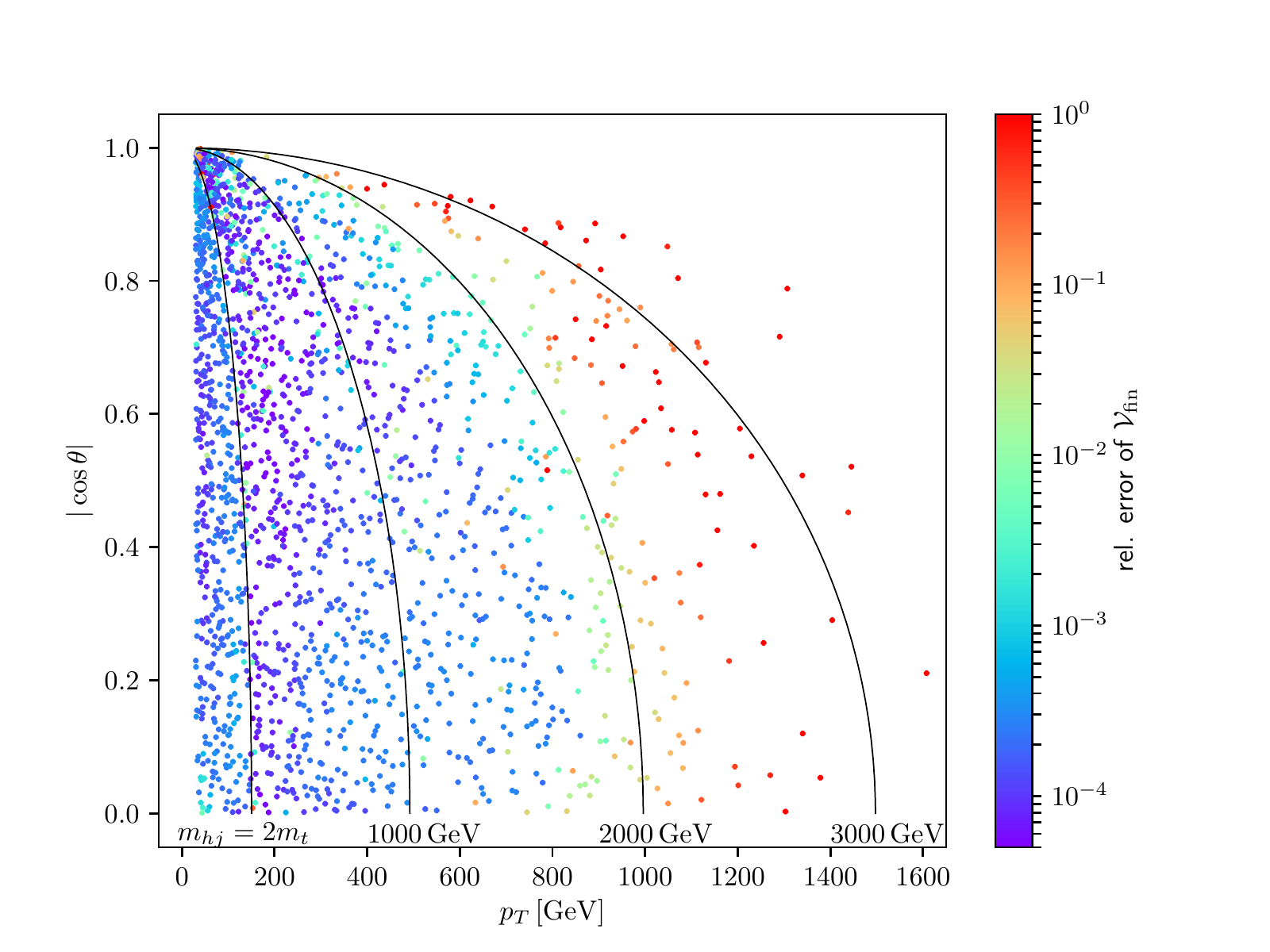}
   \includegraphics[width=0.49\textwidth]{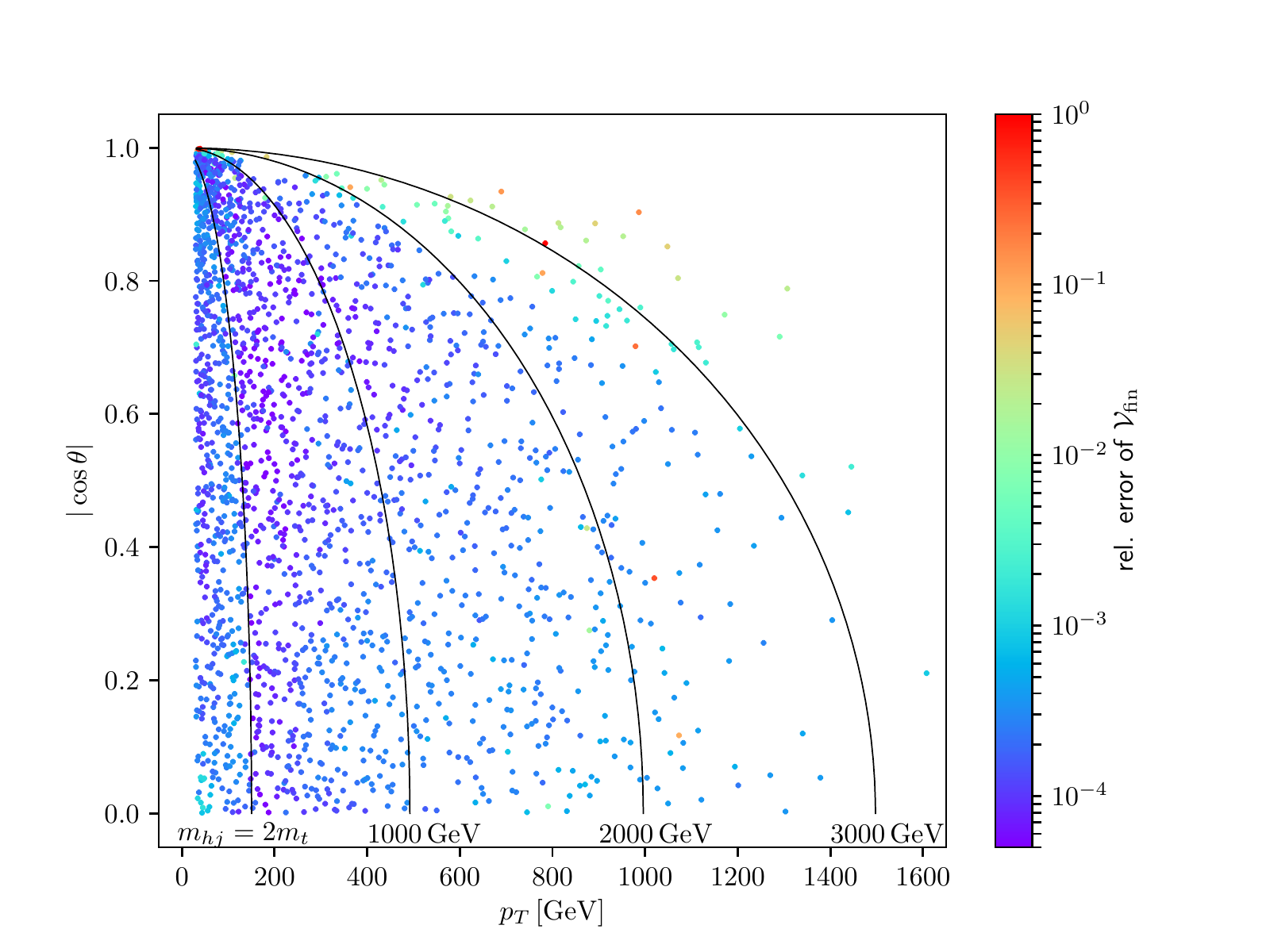}
  \caption{\label{fig:accuracy}%
    Numerical accuracy of the virtual amplitude results. The left plot shows the results used in Ref.~\cite{Jones:2018hbb}, whereas the plot on the right hand side shows results after implementing the improvements discussed in section~\ref{sec:improvements}. The finite part of the amplitude, $\mathcal{V}_\mathrm{fin}$, is defined as in Ref.~\cite{Alioli:2010xd} and parton luminosities are included, summing over all subprocesses.
    }
\end{figure}

In a numerical calculation of loop amplitudes, the choice of master integrals is very important. In particular, choosing a basis of finite integrals can be crucial for the convergence of the amplitude results. This still leaves us with an infinite space of integrals, and in Ref.~\cite{Jones:2018hbb} we have chosen the finite integrals with the least number of squared propagators and dimension shifts as master integrals. In the left plot of figure~\ref{fig:accuracy}, we  show the numerical uncertainty of the amplitude-level results obtained using this basis. For most phase space points with invariant mass $m_{hj}$ below 2\,TeV the precision goal of 0.5\% for each form factor has been reached, typically leading to amplitude results with per-mill level precision. However, at larger invariant masses, the numerical integrations converge much slower and for many phase-space points the requested precision couldn't be reached within a run-time limit of 48 GPU-hours.

Furthermore, compiling the amplitude code required multiple weeks of CPU-time, which could not be fully parallelized due to high memory usage. This was caused by the large size of the coefficients of the integrals. The corresponding source code has a combined file size of 360\,MB and requires the use of a multi-precision library to guarantee the stable evaluation of the coefficients.

To reduce these problems, we have been searching for a better basis of master integrals, aiming for the following improvements:
\begin{itemize}
  \item better convergence of the master integrals,
  \item simpler coefficients of integrals,
  \item avoid spurious poles.
\end{itemize}

One possible cause of poorly converging integrals are Landau singularities, which appear if the second Symanzik polynomial, appearing in the denominator of Feynman parametrized loop integrals, vanishes within the integration region. In the integrand functions generated by \textsc{SecDec}, these regions are avoided by deforming the integration contour, but large cancellations during the integration of the loop integral might remain. If the second Symanzik polynomial is raised to higher powers, these cancellations might be further amplified. We therefore expect a better convergence for loop integrals  with mass dimension -2, where the second Symanzik polynomial appears with exponent 1.

%
To  simplify the coefficients of the integrals, we  try to find a basis of master integrals, where the coefficients stemming from the reduction are simple. 
Since applying a basis change to the full reduction is too time consuming to test many different basis choices, we apply the  following procedure sector by sector, setting all sub-sectors to zero. After  a good basis has been found, we then apply the corresponding basis change to the amplitude using the full reduction.\\
For each sector we create a list of $\mathcal O(10)$ candidate master integrals (preferably with mass dimension -2) and we iterate over all sets of these integrals which are a valid basis. For each of these sets we obtain the basis change and apply it either to the expressions of the amplitude or, for simplicity, to some IBP relations of the corresponding sector. We then analyze the coefficients generated by the basis change and select a basis which leads to simple expressions, according to the following criteria:
\begin{itemize}
  \item denominators factorize into simple factors,
  \item dependence on space-time dimension $d$ factorizes in denominators of coefficients,
  \item avoid poles in $\varepsilon = (4-d)/2$,
  \item file size of generated expressions.
\end{itemize}
Following this procedure, the poles in $1/\varepsilon$ of the amplitude are mainly generated by integrals with a low number of propagators. This can also reduce cancellations between integrals  with a higher number of propagators.
Summing multiple rational functions is one of the most common steps during the reduction or applying a basis change.
Avoiding complicated denominator factors can therefore help keeping the resulting expressions simple.

Applying a basis change to the HJ amplitude following this procedure lead to several improvements of our  code for evaluating the virtual amplitude. The combined file size of the coefficient functions reduced from 360\,MB to 100\,MB, leading to a significant improvement in the compile time. As shown in the plot on the right-hand side of figure~\ref{fig:accuracy}, the convergence of the amplitude results also improved considerably, giving stable results for nearly all phase space points evaluated. Furthermore, the median GPU-time required for  evaluating the virtual amplitude reduced from approximately 15\,h to less than 2\,h.

\section{Results}
In this section we present results for \hj production at the LHC at a center-of-mass energy of 13 TeV. Jets are defined using the anti-kt algorithm with $R=0.4$ and $p_{t,j}>30\,\mathrm{GeV}$. We use $\overline{\mathrm{MS}}$ renormalization with 5 light quark flavors and choose the default renormalization and factorization scale $\mu_F=\mu_R=H_T/2$, with $H_T=\sqrt{\mh^2+\pth^2}+\sum_{i}\left|p_{t,i}\right|$. Scale uncertainties are estimated by the usual 7-point variation. The PDF4LHC15{\tt\_}nlo~\cite{Butterworth:2015oua,CT14,MMHT14,NNPDF}
parton distributions are used via the LHAPDF~\cite{Buckley:2014ana} interface. The masses of the Higgs boson and top quark are set to $\mh=125\,\mathrm{GeV}$ and $\mt=\mh\,\sqrt{23/12}\approx 173.055\,\mathrm{GeV}$.

 \begin{table}
     \caption{\label{tab:xsec}
     Total cross section of \hj production with the input parameters given in the text. Results with full top-mass dependence, as well as the two approximated results \heft and \ftap are shown.
     }%
   \begin{tabular}{ l  c  c }
   \hline
   \hline
   {\sc Theory} & LO [pb] & NLO [pb]\\
   \hline
   \heft: & $\sigma_{\mathrm{LO}} = 8.22^{+3.17}_{-2.15}$ & $\sigma_{\mathrm{NLO}} = 14.63^{+3.30}_{-2.54}$ \phantom{\Big|}\\
   \ftap: & $\sigma_{\mathrm{LO}} = 8.57^{+3.31}_{-2.24}$ & $\sigma_{\mathrm{NLO}} = 15.07^{+2.89}_{-2.54}$ \phantom{\Big|}\\
   Full:  & $\sigma_{\mathrm{LO}} = 8.57^{+3.31}_{-2.24}$ & $\sigma_{\mathrm{NLO}} = 16.01^{+1.59}_{-3.73}$ \phantom{\Big|}\\
   \hline
   \hline
   \end{tabular}
 \end{table}

In Table~\ref{tab:xsec} we list the total cross sections in the full theory, as well as in the \heft and \ftap approximations, where the latter one includes the full top-mass dependence in the LO and real radiation contributions, while the virtual corrections are evaluated in the \heft and rescaled by the ratio $B_{\mathrm{full}}/B_{\mathrm{HEFT}}$ of the LO matrix elements in the full theory and \heft. We find that the top-mass effects increase the NLO cross section by 9\%(6\%) relative to the NLO \heft(\ftap) result.

\begin{figure}[t!]
  \centering
   \includegraphics[width=0.49\textwidth]{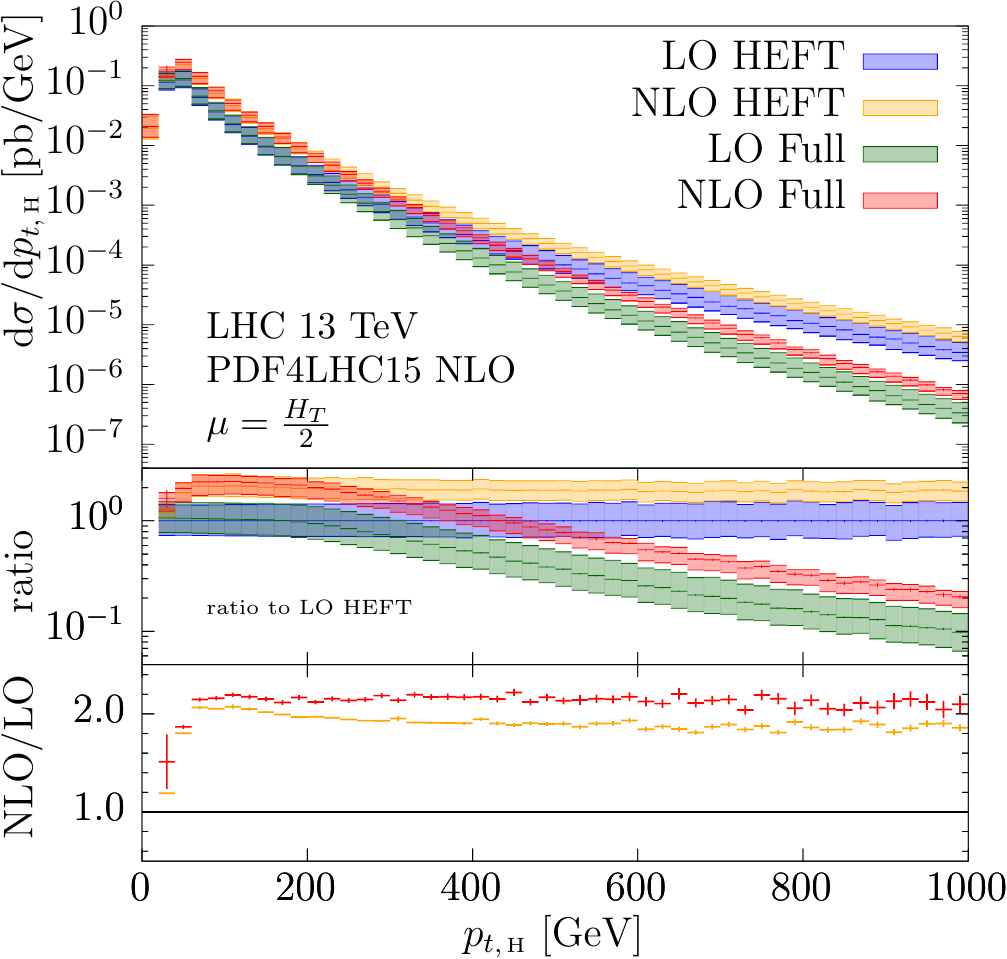}
  \includegraphics[width=0.49\textwidth]{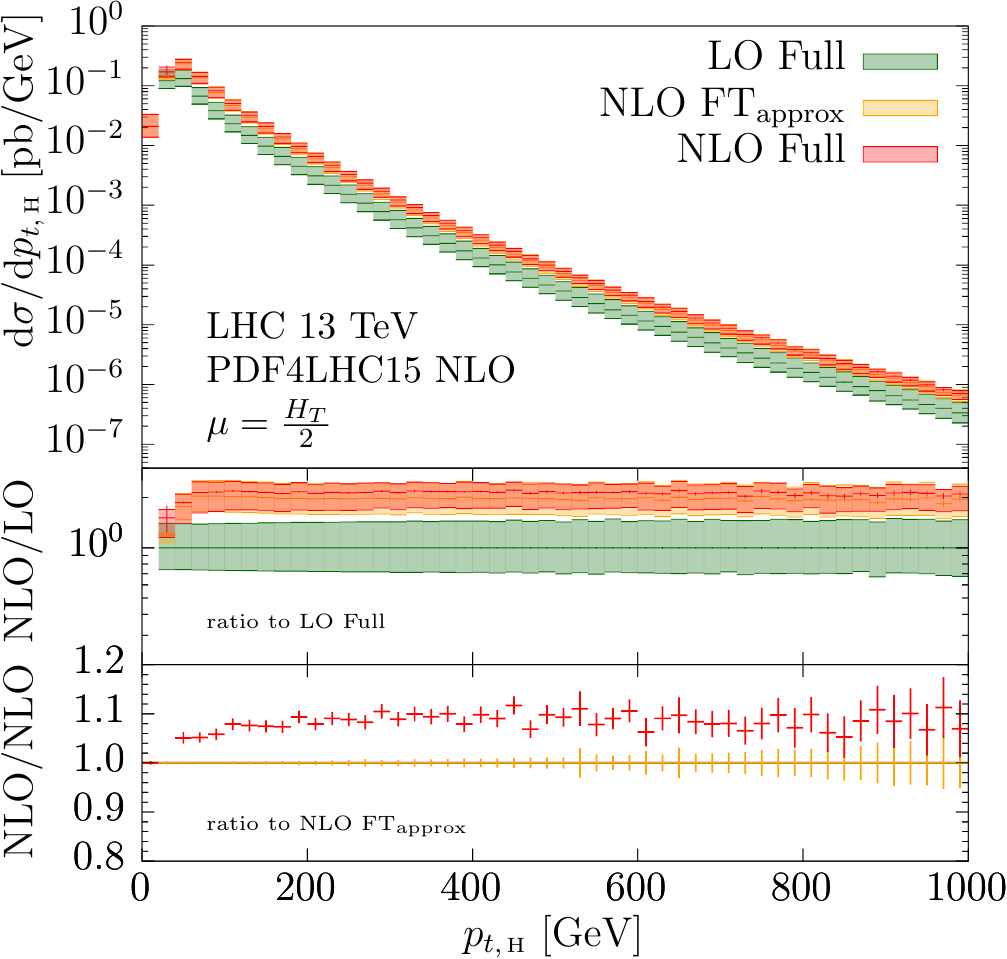}
  \caption{\label{fig:pth}%
    Higgs boson transverse momentum distribution at LO and NLO in QCD.
    The results with full top-mass dependence are compared to the \heft~(left) and \ftap~(right). The small panels show ratios of these results. Compared to Ref.~\cite{Jones:2018hbb}, the numerical uncertainties of the results are reduced due to the changes discussed in section \ref{sec:improvements}.
    }
\end{figure}  

The dependence of the cross section on the Higgs boson transverse momentum, \pth, is shown in 
Figure~\ref{fig:pth}. A comparison of the full theory and \heft result, given in the left plot, shows significant deviations of the two predictions, which can be attributed to a different scaling behavior of the amplitudes at large \pth~\cite{Forte:2015gve,Caola:2016upw}. Despite these large differences, both results lead to K-factors of similar size. However, while the K-factor is nearly constant in the full theory, it slightly decreases in the \heft al large \pth. The plot on the right-hand side shows a comparison with the \ftap result, which leads to results similar to the full theory. Taking the full top-mass dependence of the virtual contributions into account leads to an increase of about 8\% over a large range of the transverse momentum, with slightly smaller corrections at low \pth.

\begin{figure}
  \includegraphics[width=0.49\textwidth]{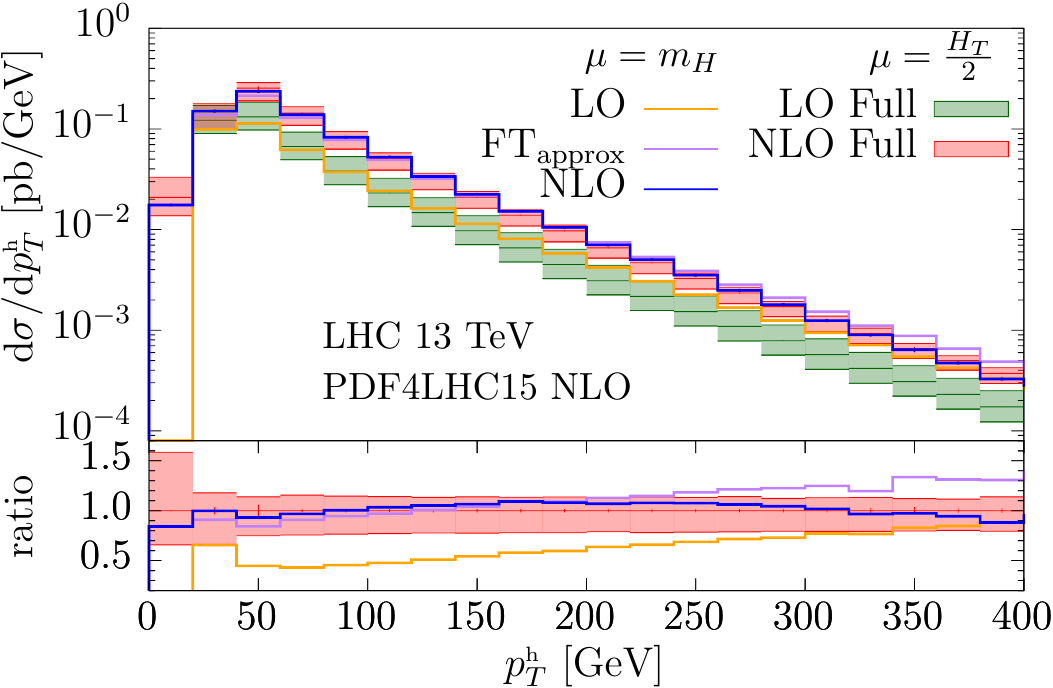}
   \caption{\label{fig:pth_mH}
     Transverse momentum spectrum of the Higgs boson using different choices of the renormalization and factorization scale. The lower panel shows the ratio to the NLO result with $\mu=H_T/2$.
     }%
 \end{figure}

Finally, in Figure~\ref{fig:pth_mH} we compare our results with the cross section evaluated using a fixed scale $\mu_R=\mu_F=\mh$. We see that, taking the full top-mass dependence into account, both NLO results are in good agreement within the scale variation band. However using a fixed scale alters the shape of the LO prediction, therefore leading to a highly phase-space dependent K-factor, which decreases from 2.1 at 50\,GeV to 1.0 at 400\,GeV. Furthermore, we obtain an enhancement in the tail of the \ftap result, which shows that retaining the full top-mass dependence leads to an  improved cancellation of the scale dependence of the NLO cross section.

\section{Conclusion}
We have presented a calculation of the NLO QCD corrections to \hj production, retaining the full dependence on the top-quark mass. Since many of the two-loop integrals appearing in the virtual corrections are not known analytically, we have evaluated all integrals numerically with the program \textsc{SecDec}. 
We have shown that the choice of the master integral basis can have a large impact on the convergence and run time of the  amplitude evaluation. Since the numerical evaluation of the virtual amplitude is very compute-intensive, 
we plan to provide our results of the virtual contribution in form of a grid interpolation framework, which will allow for a fast evaluation of these contributions without the need of numerical integrations.

The top-mass effects increase the NLO cross section of \hj production by 9\% with respect to the \heft result, but significantly reduce the cross section in the high \pth region. We also compared our results with an approximation \ftap, which reproduces the correct shape of the $p_t$-distribution, leading to differences of less than 10\% with respect to the full result in the whole $p_t$-range considered. We have also presented results using a fixed scale $\mu_R=\mu_F=m_H$, where we obtain a large phase-space dependence of the K-factor and larger top-quark mass effects, which are required to obtain good agreement with the NLO result using a dynamical scale.

\section*{Acknowledgments}
I  thank S.~Jones and G.~Luisoni for the very good collaboration and for comments on this manuscript.
I want thank Stephan Jahn and Lorenzo Tancredi for useful discussions about choosing a good basis of  master integrals. We gratefully acknowledge
support and resources provided by the Max Planck Computing and Data
Facility (MPCDF).

\bibliography{references}

\end{document}